\documentstyle[epsf]{mn}
\begin{document}
\renewcommand{\baselinestretch}{1}
\newcommand{\bnabla}{{\boldsymbol{\nabla}}}
\newcommand{\bu}{\mathbf{u}}
\newcommand{\bB}{\mathbf{B}} \newcommand{\be}{\mathbf{e}}
\def\msun{{\rm M}_\odot}
\def\rsun{{\rm R}_\odot}
\def\halpha{H$\alpha$}
\def\hbeta{H$\beta$}
\newcommand{\lsimeq}{\mbox{$\, \stackrel{\scriptstyle <}{\scriptstyle \sim}\,$}}
\newcommand{\gsimeq}{\mbox{$\, \stackrel{\scriptstyle >}{\scriptstyle \sim}\,$}}
\def\gradi{\ifmmode{^\circ}\else$^\circ$\fi}
\def\reference{\parskip 0pt\par\noindent\hangindent 0.5 truecm}
\def\kms{km ${\rm s}^{-1}$} 
\def\Msun{{\rm M}_\odot}

\title[Birth properties of MSPs]{The birth properties of Galactic millisecond 
radio pulsars}

\author[L. Ferrario \&  D. T. Wickramasinghe]
{Lilia Ferrario and Dayal Wickramasinghe \\
Department of Mathematics, The Australian National University,
Canberra, ACT 0200, Australia}

\date{Accepted.  Received ; in original form} 

\maketitle

\begin{abstract}

We model the population characteristics of the sample of millisecond pulsars
(MSPs) within a distance of $1.5$~kpc.  We find that for a braking index
$n=3$, the birth magnetic field distribution of the neutron stars as they
switch on as radio-emitting MSPs can be represented by a Gaussian in the
logarithm with mean $\log B(G)= 8.1$ and $\sigma_{\log B}=0.4$ and their birth
spin period by a Gaussian with mean $P_0=4$~ms and $\sigma_{P_0}=1.3$~ms. We
assume no field decay during the lifetime of MSPs. Our study, which takes into
consideration acceleration effects on the observed spin-down rate, shows that
most MSPs are born with periods that are close to the currently observed
values and with average characteristic ages that are typically larger by a
factor $\sim 1.5$ compared to the true age.  The Galactic birth rate of the
MSPs is deduced to be $\gsimeq 3.2 \times 10^{-6}$~yr$^{-1}$ near the upper
end of previous estimates and larger than the semi-empirical birth rate $\sim
10^{-7}$~yr$^{-1}$ of the Low Mass X-ray Binaries (LMXBs), the currently
favoured progenitors. The mean birth spin period deduced by us for the radio
MSPs is a factor $\sim 2$ higher than the mean spin period observed for the
accretion and nuclear powered X-ray pulsars, although this discrepancy can be
resolved if we use a braking index $n=5$, the value appropriate to spin down
caused by angular momentum losses by gravitational radiation or magnetic
multipolar radiation.  We discuss the arguments for and against the hypothesis
that accretion induced collapse (AIC) may constitute the main route to the
formation of the MSPs, pointing out that on the AIC scenario the low magnetic
fields of the MSPs may simply reflect the field distribution in isolated
magnetic white dwarfs which has recently been shown to be bi-modal with a
dominant component that is likely to peak at fields below $10^3$~G which would
scale to neutron star fields below $10^9$~G, under magnetic flux conservation.

\end{abstract}


\section {Introduction}

The properties of the MSPs and the ``normal'' radio pulsars place them in two
nearly disjoint regions in the spin period ($P$) period derivative ($\dot P$)
diagram.  In the normal pulsars, $P$ and $\dot P$ are distributed about mean
values of $\sim 0.6$~s and $\sim 10^{-15}$~s~s$^{-1}$ respectively, with
implied magnetic field strengths in the range $\sim 10^{11}-10^{13}$~G.  In
contrast, in the MSPs, $P$ and $\dot P$ are distributed about $\sim 5$~ms and
$\sim 10^{-20}$~s~s$^{-1}$ respectively, with field strengths in the range
$\sim 10^8-10^9$~ G. This bi-modality in the field distributions of the radio
pulsars has been an enigma which has still to be fully resolved.

There are also major differences in the population characteristics of these
two groups of pulsars which provide important clues on their origin. Most
($\sim 85$\%) of the MSPs are in binary systems (MSPs) on nearly circular
orbits in contrast to the normal pulsars which tend to be isolated and when
they are not, exhibit more eccentric orbits. Furthermore, proper motion
studies have shown that while the average space velocity for normal pulsars is
$\sim 400$~km~s$^{-1}$ (Hobbs et al. 2005), the MSPs form a low-velocity
population with typical transverse speeds of $\sim 85$~km s$^{-1}$ (Hobbs et
al. 2005; Toscano et al. 1999).  Differences in the incidence of binarity,
eccentricity of orbits, and space motions are usually attributed to
differences in the kick velocity imparted to the neutron stars at birth
(Shklovskii 1970). The magnitude of the kick, and its effect on the binary
system, will depend on the nature of the system, and whether the neutron star
originates from the core collapse of a massive star with a supernova
explosion, or from the accretion induced collapse (AIC) of a white dwarf.

In the standard model, the MSPs are considered to be the end product of the
evolution of low-mass and intermediate-mass X-ray binaries where it is assumed
that the neutron star was formed by core collapse (CC) of a massive ($M>
8\Msun$) star, and is subsequently spun up to millisecond periods during an
accretion disc phase (Bhattacharya \& van den Heuvel 1991; Bisnovatyi-Kogan \&
Komberg 1974). We shall refer to this class of objects as the ``Core-Collapsed
LMXBs and IMXBs'' or, more briefly, the LMXBs(CC)/IMXBs(CC). Accretion induced
field decay is an integral part of this model which appears plausible from a
theoretical view-point, particularly if the fields in neutron stars are of
crustal origin (Konar \& Bhattacharya 1997, but see Ruderman 2006 for an
alternative model).  Regardless of the origin of the low fields in the MSPs, a
long standing problem with the LMXB/IMXB scenario has been the difficulty in
reconciling their semi-empirical birth rates with those of the radio MSPs
(e.g. Lorimer 1995; Cordes \& Chernoff 1997).  The problem with the birth
rates has been confirmed by recent population synthesis calculations which
have also highlighted the difficulties in explaining the observed orbital
period distribution of MSPs on the LMXB(CC)/IMXB(CC) scenario (Pfhal et
al. 2003).

Another often discussed channel for the production of MSPs involves the AIC of
an ONeMG white dwarf (Michel 1987). Here, during the course of mass transfer,
a white dwarf reaches the Chandrasekhar limit, and collapses to form a neutron
star (Bhattacharya \& Van den Heuvel 1991).  In the AIC scenario, we may
expect the magnetic field distribution of the MSPs to reflect in some way the
magnetic field distribution of their progenitor white dwarfs obviating the
need for field decay. A low-mass or intermediate-mass X-ray binary phase may
follow the collapse of the white dwarf and we shall refer to this class of
objects as the ``Accretion Induced Collapse LMXBs and IMXBs'' or, more
briefly, as the LMXBs(AIC)/IMXBs(AIC). Population synthesis calculations
indicate that the expected birth rates from the AIC channel may be
significantly higher than those from the LMXB(CC)/IMXB(CC) route (Hurley et
al. 2002; Tout et al 2007; Hurley 2006, private communication).

In this paper, we present an analysis of the 1.5~kpc sample of MSPs which is
considered to be sufficiently sampled out (Lyne et al. 1998; Kramer et
al. 1998) with the aim of establishing the MSP birth properties and
constraining the different models that have been proposed for their origin.
Our estimate of the Galactic birth rate of the MSPs is at the upper end of
previous estimates (e.g. Cordes \& Chernoff 1997) and again brings into
question the LMXB(CC)/IMXB(CC) scenario as being the dominant route for the
origin of the MSPs.  The paper is arranged as follows. In section 2 we
describe the data set and our model.  Our results are presented and discussed
in section 3 where we also present the case for and against LMXB(CC)/IMXB(CC)
progenitors and AIC progenitors for the MSPs.  Our conclusions are presented
in section 4.

\section{The modelling of the radio properties of the MSPs}

In 1998, Kramer et al. conducted a very detailed study aimed at comparing the
radio emission properties of MSPs to those of normal pulsars. In their work,
they restricted their comparative studies to objects within 1.5 kpc, on the
grounds that the population of all radio-pulsars is sufficiently sampled out
up to this distance, as first pointed out by Lyne et al. (1998). Recently,
this assumption has gained further strength through the high-latitude survey
of Burgay et al. (2006) in the region of the sky limited by
$220\gradi<l<260\gradi$ and $|b|<60\gradi$ conducted with the 20-cm multi-beam
receiver on the Parkes radio-telescope.  If we restrict the pulsars in this
survey to a distance of up to 1.5~Kpc, we find that all but one previously
known radio-pulsars have been re-detected and 4 new objects, out of a total of
16, discovered. We therefore apply a correction factor of $1.25$ to obtain an
estimate of the total number of MSPs in the $1.5$ kpc sample that we analyse,
stressing the fact that this may only be a lower limit for the real number of
objects up to this distance.

We have restricted our analysis to binaries and isolated millisecond pulsars
with spin periods shorter than 30 ms. In the restricted sample to 1.5 kpc, the
Australia Telescope National Facility (ATNF) catalogue (Manchester et al. 2005) gives 24 MSPs in binaries and 11
isolated MSPs. We do not distinguish between these two groups since the
isolated MSPs are likely to be an end product of binary systems in which the
companion has been tidally disrupted or ablated (Radhakrishnan \& Shukre 1986)
with an otherwise similar evolutionary history to the binary MSPs (see,
however, Michel 1987; Bailyn \& Grindlay 1990 for alternative points of
view). They exhibit very similar observational properties, except, perhaps,
for the radio-luminosity which seems to be slightly lower in the isolated MSPs
(Bailes et al. 1997). Of the $24$ binary MSPs, $14$ have periods above $10$~d
and most have lower limits to the companion masses in the range $\sim 0.2 -
0.4 \Msun$ (Manchester et al. 2005) indicative of the end state of evolution
of binary systems that evolve to longer periods (beyond the bifurcation
period, see Podsiadlowski, Rappaport \& Pfahl 2002) due to mass transfer from
a low-mass giant (see sections 3.1 and 3.2) leading to He white dwarfs. The
remaining shorter period systems appear to have either He or CO white dwarfs
or very low-mass companions.

The age of the millisecond pulsar is calculated according to
\begin{equation} 
\label{age}
t=\frac{P}{(n-1)\dot P}\left[1-\left(\frac{P_0}{P}\right)^{n-1}\right]
\end{equation}
where $n$ is the braking index, which is equal to 3 for the dipolar spin-down
model, and $P$ and $P_0$ are the observed and the initial period of the MSP
respectively. If $P_0<<P$ we obtain the characteristic age of the MSP:
\begin{equation}
\label{char_age}
\tau_c=\frac{P}{2\dot P}.
\end{equation}
In the next sections we will show that $P_0$ is often too close to $P$ to be
able to rely on $\tau_c$ for an estimate of the true age of a MSP.

The very low spin-down rates of the MSPs have so far precluded any direct
measurements of the braking index. An index of $n=3$ is indicated for old
normal radio pulsars, but values $n\sim 1.5-2.8$ have been measured in younger
pulsars (Lyne 1996; Hobbs et al. 2004). In this work, we have adopted
$n=3$, but it is conceivable that a different value of $n$ may be appropriate
for the MSPs. For instance, if spin down is by multi-polar radiation, the
braking index will be somewhat larger than $3$, while if angular momentum is
lost mainly by gravitational radiation, we expect $n=5$ (Camilo, Thorsett \&
Kulkarni 1994). In our modelling we adopt $n=3$ but we also discuss the
implications of using a larger value of $n$.

We synthesise the properties of the MSPs using essentially the method
described in Ferrario \& Wickramasinghe (2006, hereafter FW). However, in the
present study there are two main differences.  Firstly, we directly assume an
initial magnetic field distribution for the MSPs without attempting to relate
it back to the magnetic properties of the (main sequence) progenitors.  We
therefore have as our basic input the MSP birth magnetic field distribution,
which we describe by a Gaussian in the logarithm, and the birth spin
distribution also described by a Gaussian. We stress that here with ``birth''
characteristics of MSPs we refer to those characteristics that the MSPs have as
they switch on as radio emitters, regardless of their previous history. Hence,
the results of our calculations do not depend in any ways on the specific
route(s) leading to the formation of the MSPs.

Secondly, we take into consideration the three Doppler accelerations effects
cited by Damour \& Taylor (1991) which affect the observed spin-down rate of
the MSPs, namely, (i) the Galactic differential rotation, (ii) the vertical
acceleration $K_z$ in the Galactic potential and (iii) the intrinsic
transverse velocity of the pulsar. Thus, the observed spin-down rate is given
by (e.g. Toscano et al. 1999)
\begin{equation}
\label{shk}
\dot P_{\rm obs} =\dot P_i + \Delta \dot P
\end{equation}
where $\dot P_i$ is the ``intrinsic'' spin-down rate and $\Delta\dot P$ is the
term due to the aforementioned acceleration effects.  Hence, when we compare
our models to observations, we introduce these acceleration terms to our
synthetic population to mimic the behaviour of the observed MSPs.

We follow the motions of the stars we generate by integrating the equations of
motion in the Galactic potential of Kuijken \& Gilmore (1989) assuming that
the neutron stars are born with a kick velocity given by a Gaussian
distribution with velocity dispersion $\sigma_v$.

To fit the observations, we also model the radio luminosity at 1400 MHZ and
compare it to the members of our list with a measured value at this
frequency. The studies of Kramer et al. (1998) and Kuz'min (2002) indicate
that despite the large differences in periods and magnetic fields, normal
pulsars and MSPs exhibit the same flux density spectra, therefore pointing
towards the same emission mechanism, although the MSPs tend to be weaker
sources on average (Kramer et al. 1998). Hence, similarly to many previous
investigators (e.g.  FW; Narayan \& Ostriker 1990), we have assumed that the
luminosity $L_{400}$ at 400~MHZ can be described by a mean luminosity of the
form
\begin{equation}
\label{lm}
\log \langle L_{400}\rangle=\frac{1}{3}\log\left(\frac{\dot
  P}{P^3}\right)+\log L_0
\end{equation}
Here the luminosities are in units of mJy~kpc$^2$. We have modelled the spread
around $L_{400}$ using the dithering function of Narayan \& Ostriker (1990) to
take into account the various intrinsic physical variations within the sources
and also variations caused by different viewing geometries. This function is
given by
\begin{equation} 
\label{gamfun}
\rho_L(\lambda)=0.5\lambda^2\exp\left(-\lambda\right)\qquad\qquad (\lambda\ge 0)  
\end{equation}
where
\begin{equation}
\label{dith}
\lambda=b\left(\log\frac{L_{400}}{\langle L_{400}\rangle}+a\right)
\end{equation}
and $a$ and $b$ are constants to be determined (Hartman et al. 1997). 

Kramer et al. (1998) find that by restricting their comparison analysis of
normal radio-pulsars to MSPs to sources up to 1.5 kpc, the mean spectral
indices of normal radio-pulsars and MSPs are essentially the same, i.e.,
$-1.6\pm 0.2$ (MSPs) and $-1.7\pm 0.1$ (normal pulsars). Hence our deduced
radio luminosity at 400 MHZ is scaled to 1400 MHZ using a spectral index
of $-1.7$ (as in FW).

Once all the intrinsic properties of our model MSPs are determined, we check
for pulsars detectability at 1400 MHZ by the Parkes multi-beam receiver
(e.g. Manchester et al. 2001; Vranesevic et al. 2004).

Furthermore, pulsars radio emission is anisotropic with pulsars at shorter
periods exhibiting wider beams, hence we need to correct for this factor,
since this will influence the birth rates of MSPs. For example, large beams would
require smaller birth rates, since the MSPs would have a greater chance to be
detected. However, there is as yet no agreement on the beaming fraction-period
relationship, particularly for the MSPs. Rankin (1993), Gil et al. (1993) and
Kramer et al. (1994) pointed out that observational evidence seems to suggest
that the opening angles of normal radio-pulsars (that is, the last open
dipolar field line) is proportional to $1/\sqrt{P}$. In the absence of a
consensus on this issue, we use Kramer's (1994) model at a frequency of 1.4
GHz for the opening half-angle $\theta$ (in degrees) of the pulsar beam:
\begin{equation}
\label{op_angle}
\theta=\frac{5.3\gradi}{P^{0.45}}.
\end{equation}
These values of $\theta$ yield duty cycles of less than unity for periods down
to about 1~ms. However, we would like to remark that our results are quite
insensitive to slight modifications to the above $\theta-P$
relationship. Then, by assuming that the viewing angles of MSPs are randomly
distributed, the fraction $f$ of the sky swept by the radiation beam is given
by (Emmering \& Chevalier 1989)
\begin{equation}
\label{beam}
f=\left(1-\cos\theta_r\right)+\left(\frac{\pi}{2}-\theta_r\right)\sin\theta_r
\end{equation}
where $\theta_r$ is the half-opening angle now in radians. We will use this $f$
to compare our MSP synthetic population to the data sample under consideration.

\section{Results and discussion}

We have used as our observational constraints the 1-D projections of the data
comprising the number distributions in period $P$, magnetic field $B$,
period derivative $\dot P$, radio luminosity $L_{1400}$,
$Z$-distribution and characteristic age $\tau_c$, as determined from equation
\ref{shk}.
\begin{figure*}
\begin{center}
\hspace{0.1in}
\epsfxsize=0.6\textwidth
\epsfbox{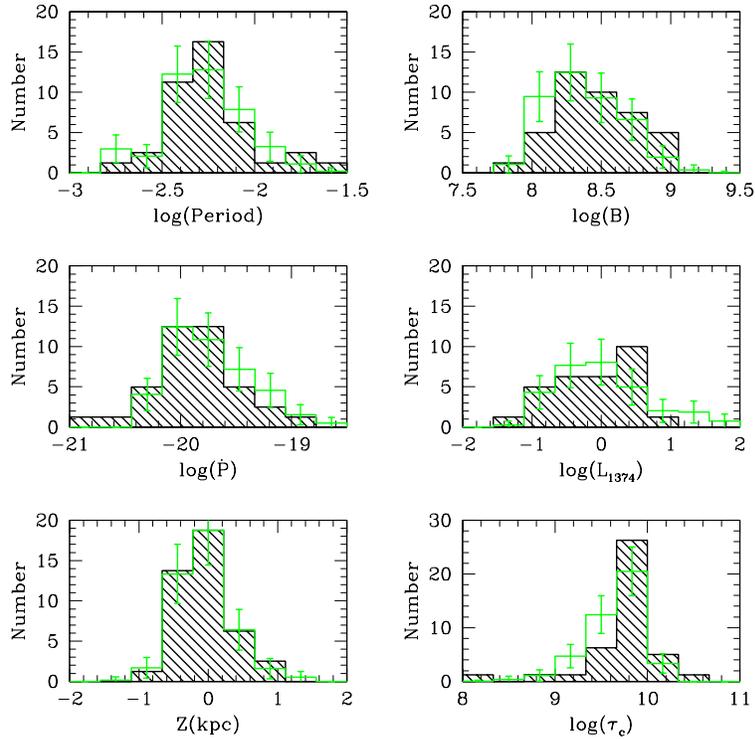}
\caption{Our model (solid line) overlapped to the MSPs data sample of up to
  1.5~Kpc (shaded) taken from the ATNF catalogue
  (http://www.atnf.csiro.au/research/pulsar/psrcat, Manchester et
  al. 2005). The error bars indicate the Poisson standard errors.}
\end{center}
\end{figure*}

Similarly to FW, the best fit model to the observations of the MSPs was
determined ``by eye'' after conducting hundreds of trials. Our results are
shown in Figure 1. The fit was obtained by setting $\sigma_{P_0}=1.3$~ms about
a mean $P_0=4$~ms. The parameters for the luminosity model are ${\langle
L_{400}\rangle}=5.4$, $a=1.5$ and $b=3.0$. Furthermore, the MSPs are imparted
with a one-dimensional natal kick dispersion velocity of $50$~km~s$^{-1}$,
which yields an average transverse velocity of $83$~km~s$^{-1}$. Our model
reproduces the observed total number of MSPs with a local formation rate of
$\gsimeq 4.5\times 10^{-9}$~yr$^{-1}~$kpc$^{-2}$ which translates into a
Galactic birth rate of $\gsimeq 3.2\times 10^{-6}$~yr$^{-1}$. We emphasise
that this should be seen as a lower limit for the MSP birth rate, since the 1.5
kpc sample is still likely to be somewhat incomplete even after the correction
factor that we have applied.

Our calculations show that the observed MSP magnetic field distribution can be
modelled with a field which is initially (that is, at the onset of the MSP
radio-emission phase) Gaussian in the logarithm. We find that, similarly to
our modelling of normal isolated radio-pulsars (see FW), it is not necessary
to assume any spontaneous field decay during the lifetime of the
radio-emitting MSPs. Thus, the ``high'' magnetic field tail of MSPs arises
from the dependence of the spin-down rate on the magnetic field, and not from
a complex birth field distribution assigned to the parent population (e.g. as
a result of accretion-induced field decay during a previous phase of mass
accretion). In this context, we note that if the MSPs were born with higher
field strengths than postulated by us and then decayed during their
radio-emission lifetime towards weaker magnetic fields values, then we would
expect a continuous field distribution filling up the gap between the normal
radio-pulsars and the MSPs. This was first noted by Camilo et al. (1994), who
also point out that there is no indication that old globular cluster MSPs have
magnetic fields which are lower than those of MSPs in the Galactic disc.

Another observational peculiarity of MSPs is that some of these objects appear
to be older than the Galactic disc (10~Gyr) if one uses equation
\ref{char_age} to assign them an age. Toscano et al. (1999) found that by
correcting their spin-down rates for the transverse velocity (Shklovskii)
effect they exacerbated this paradox, since the correction resulted in a
decrease in the spin-down rate $\dot P$ (and thus of the derived magnetic
field strength) and an increase in characteristic age. As a consequence,
nearly half of their corrected sample exhibited characteristic ages comparable
to or greater than the age of the Galactic disc. Our theoretical results agree
with their findings and are presented in Figure 2 where we have plotted the
observed $\dot P_{obs}$ and intrinsic $\dot P_i$ spin-down rates of our
synthetic population. This clearly supports the view that transverse velocity
effects do play an important role in the observations of MSPs.
\begin{figure}
\begin{center}
\hspace{0.1in}
\epsfxsize=0.75\columnwidth
\epsfbox{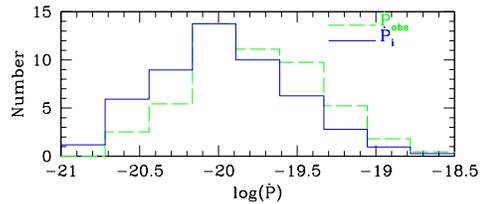}
\caption{Intrinsic (solid histogram) and observed (dashed histogram) spin-down 
rates of our synthetic population of MSPs.}
\end{center}
\end{figure}

Camilo et al. (1994) proposed three possibilities to solve this paradox.  The
first is that the magnetic field structure of MSPs is multipolar and thus the
braking index that appears in equation \ref{age} may be greater than
3. Alternatively, due to gravitational radiation, $n=5$. The second is that
the magnetic field decays, so that high values of $\tau_c$ could be attained
by a certain choice for the decay time-scale, although they discarded this
possibility as outlined earlier. Finally, the third possibility that Camilo et
al. (1994) proposed is that at least some MSPs may have been born with
$P_0\sim P$.

Our modelling has shown that if we start with an initial period distribution
that is Gaussian with mean at 4~ms, we obtain a current day period
distribution that is close to what is observed. In particular, we can
reproduce the sharp rise in the observed period distribution near $\sim 3$
ms. In our model, most pulsars have initial periods that are quite close to
the observed periods. This yields an average characteristic age which is
larger than the average true age (as given in equation \ref{age}) by nearly
50\%.  Further observational support in favour of birth periods being close to
observed periods in MSPs also comes from studies of individual systems.  For
instance, for PSR J0437-4715, Johnston et al. (1993) and Bell et al. (1995)
find $P=5.757$~ms, $P_{\rm orb}=5.741$~d and $\tau_c=4.4-4.91$. Sarna, Ergma,
Ger\u{s}kevit\u{s}-Antipova (2000) derive the mass of the companion of PSR
J0437-4715 to be $0.21\pm 0.01$~M$_\odot$ with a cooling age of
$1.26-2.25$~Gyr. This implies that PSR J0437-4715 is much younger than
inferred through its characteristic age and was born with a period close to
the current period.

The lack of sub-millisecond pulsars is apparent in the distribution of the
MSPs and has also been noted in the millisecond X-ray pulsars (Chakrabarty
2005).  Our study indicates that this is consistent with our assumption of a
Gaussian distribution of initial birth periods that peaks at $4$~ms and rules
out the possibility that most millisecond pulsars are born at sub-millisecond
periods. Here, selection effects may be playing a role, however, Camilo et
al. (2000) estimated a loss of sensitivity of only 20\% below about 2~ms,
which is far too low to explain the sudden drop in the number of MSPs below
this period. This may suggest that neutron stars can never achieve break-up
spin periods ($\sim 0.4-0.7$~ms depending on equation of state, Cook et
al. 1994). Hence, loss of angular momentum caused by gravitational radiation
may limit the neutron star spin rate (Wagoner 1984). Current estimates to the
lower limits for spin periods are about 1.4~ms (Levin \& Ushomirsky 2001),
which is close to the spin of the recently discovered MSP in the globular
cluster Terzan 5 (Hessels et al. 2006).

\subsection{The LMXB(CC)/IMXB(CC) scenario and its relation to birth properties}

In the LMXB(CC) route, matter is transfered from low-mass main-sequence
donors and binary evolution occurs towards shorter periods driven by magnetic
braking and gravitational radiation. Mass transfer continues past the period
minimum over a Hubble time or until the companion is evaporated.  Binary
MSPs that result from this route are expected to have very low-mass
companions with ultra short orbital periods. In contrast, in the IMXB(CC)
route the donor stars are of intermediate mass $\gsimeq 2\Msun$, and binary
evolution is driven by nuclear evolution past the bifurcation period towards
longer periods.  For the lower mass donor stars, mass transfer phase ends when
the helium core of the donor star is exposed as a low-mass ($\sim 0.2- 0.4
\Msun $) helium white dwarf. For the more massive donor stars, mass transfer
can terminate when a CO or an ONeMg white dwarf core is exposed.  Although the
observed sample of binary MSPs consists of systems with all of the above
companions, recent population synthesis calculations have not been successful
in modelling the observed orbital period distributions (Pfahl et al. 2003).

Thus, Pfahl et al. (2003) find that the LMXB(CC) route leads to a significant
population of low period binary MSPs peaking at $P_{\rm orb}\sim 0.03$~d, but this
population is not represented in the observed sample of MSPs.  Indeed, the
shortest observed binary period for the MSPs is $P_{\rm orb} = 0.1$~d for
PSR~J2051-0827 (Stappers et al. 2001). On the other hand, the binary periods
of the ultra-compact X-ray binaries are generally significantly shorter with a
few of them exhibiting orbital periods of 0.03 days. Hence, this may be an
indication that either (i) as their LXMB(CC) evolution continues, they will
end up ablating their companion and thus appearing as isolated MSPs, or (ii)
the ultra-compact LMXBs(CC) and the MSPs are not evolutionarly linked.

In contrast, the IMXB(CC) route predicts a population of binaries that peaks
roughly at the observed periods $P_{\rm orb} \sim 6-30$~d, but with a width
that falls short by a factor of $\sim 10-100$ (depending on assumptions on the
common envelope parameter) in comparison to the observations of binary MSPs.
Indeed, the majority of the binary MSPs in the ATNF sample (Manchester et
al. 2005) do not have orbital periods that fall in the most probable region
(Pfahl et al. 2003) predicted for either LMXB(CC) or IMXB(CC) evolution.

There is also the problem with the birth rates mentioned in section
1. Attempts at reconciling the LMXB(CC)/IMXB(CC) rates with the birth rates of
MSPs have not been successful (Pfahl et al. 2003) and the present results go
in the direction of making this discrepancy larger. It has been suggested that
the above discrepancies may disappear when more realistic models are
constructed that allow for limit cycles that may arise from X-ray irradiation
of the donor stars, and for the intricacies in common envelope evolution. This
remains a possibility.

However, it should be noted that the semi-empirical birth rates are based
almost entirely on the observed LMXBs, which, given the arguments above,
cannot be the dominant progenitors of the binary MSPs. The same comment also
applies to our discussion of the AIC that undergo a phase of mass transfer
after collapse (see section 3.2).

In the evolution that leads up to LMXBs(CC) and IMXBs(CC), one of the stellar
components (usually, the initially more massive) evolves into a neutron star
through core collapse with a field distribution peaking near $\log
B(G)=12.5$ (as observed in the isolated radio-pulsars). The observed field
distribution of the MSPs, on the other hand, peaks at $\log B(G) = 8.4$.  This
discrepancy is often explained by accretion-induced field decay or
evolution. The presence of MSPs in old systems suggests that if the low value
of the magnetic field is due to field decay, it must do so mainly during the
accretion phase prior to the neutron star becoming a radio MSP.

The manner in which the field is expected to decay in accreting neutron stars
depends on the origin of the magnetic fields, and here there is no
consensus. There is no clear evidence for field decay in ordinary pulsars on a
time scale of $10^7 - 10^8$~yr.  However, accretion can enhance field decay,
particularly if the fields are of crustal origin. Two competing effects have
been considered.  Accretion raises the temperature and reduces the
conductivity in regions of the crust that carry the current, thereby enhancing
field decay. Accretion also pushes the current forming region inward towards
regions of higher density and conductivity where the field can be frozen.
Konar \& Bhattacharya (1997) have shown that these two effects, when taken
together, could lead to an asymptotic ``frozen'' field strength that is a
factor of $10^{-1}-10^{-4}$ below the initial field strength. The asymptotic
value depends on the accretion rate and the total mass accreted. On the other
hand, Wijers (1997) presented strong evidence against accretion-induced field
decay which is proportional to the accreted mass onto the neutron star. Thus,
the standard model does not explain the observed characteristics of the
MSP birth field distribution as they switch-on as radio-emitters.  For
instance, if we consider the route that contributes to the majority of binary
MSPs, namely those having orbital periods $P_{\rm orb} \ge 10 -1000$~d with
low mass He WD companions arising from the evolution of intermediate mass
donors, we may expect the accretion history, and therefore the birth field
distribution, to depend on the orbital period.  It is therefore not
immediately apparent why this field should have a nearly Gaussian distribution
with such a narrow width.  The problem becomes even more severe when more than
one channel is considered (see discussion in Tout et al. 2007).

The detection of coherent X-ray pulsations with millisecond periods in a
handful of LMXBs (Lamb \& Yu 2005) is often used in support of the idea of
accretion induced field decay (Wijnands \& van der Kliss 1998).  However,
whether this is evidence simply for field submersion and spin up during an
accretion disc phase, or for field decay and spin up, remains to be
established. Cumming et al. (2001) have argued that the majority of the LMXBs
do not show coherent pulsations because they may have fields significantly
less than $10^8$~G due to field submersion which, at face value, is
inconsistent with the fields seen in the radio-MSPs, but their calculations
also indicate that the field will re-emerge on a time scale of $\sim 1000$~yr
although it is unclear to what value. Indeed, for the LXMB(CC)/IMXB(CC)
standard scenario to be viable, the field would be required to re-emerge to
values that are similar to those observed in the radio MSPs.

If we adopt the contentious viewpoint that magnetic fields do not decay due to
accretion, but are simply temporarily submerged, and re-emerge to their
original values of a few $\times 10^{12}$~G at the end of the
LMXB(CC)/IMXB(CC) phase, then we may expect a population of high field MSPs.
The objects in such a population would have a birth rate that is $10^{-4}$
times the birth rate of normal radio-pulsars and would therefore be unlikely
to be represented in the current sample of radio-pulsars. Furthermore, since
they would spin down very rapidly to much longer periods (with characteristic
time scales of only a few hundred years), they would have an even smaller
chance to be detected as high field radio-MSPs.

Finally, we note that on the LMXB(CC)/IMXB(CC) hypothesis, we expect the birth
spin period distribution of the MSPs as they become radio-emitters, to be
similar to the observed spin period distribution of the LMXBs(CC). However,
observations of accretion and nuclear powered LMXBs show that their spin
periods peak near 2~ms (Lamb \& Yu 2005). This could indicate either a
different origin for the radio MSPs (see section 3.2), or that the braking
index is significantly larger than adopted by us.  We have carried out
calculations for different braking indices and find that a braking index of
$n= 5$ (appropriate to angular momentum loss by gravitational radiation or
magnetic multipolar radiation) will bring the two distributions into closer
agreement.

\subsection{The AIC scenario and its relation to birth properties}

According to current models, accretion induced collapse leads to the formation
of a rapidly spinning (a few milliseconds) neutron star when an ONeMg white
dwarf accretes matter in a binary system and reaches the Chandrasekhar limit.
Recent calculations have re-affirmed that an AIC is the expected outcome of
thermal time scale mass transfer in such systems with orbital periods of the
order of a few days (Ivanova \& Taam 2004).  The magnetic fluxes in these
cores may thus reflect the magnetic fluxes seen in the isolated white dwarfs.

Until recently, it was believed that the magnetic fields of the isolated white
dwarfs could be described by a single distribution.  However, it is now
evident that the distribution is bi-modal, comprising of a high and a low
field component. The high field component ($10-15$\% of all white dwarfs) has
a distribution that peaks at $\log B(G) \sim 7.5 $ with a half width
$\sigma\log B = 7.3$ (e.g. Wickramasinghe \& Ferrario 2005).  This
distribution declines towards lower fields with very few stars detected in the
field range $10^5 - 10^6$~G (magnetic field gap). The incidence of magnetism
rises again towards lower magnetic fields with some $15 - 25$\% of white
dwarfs being magnetic at the kilo-Gauss level (Jordan et al. 2006). Since the
new detections of Jordan et al. (2006) are at the limit of the sensitivity of
current spectropolarimetric surveys, it appears likely that \emph{all} white
dwarfs will be found to be magnetic, with the majority ($\sim 85$\%) belonging
to the low field group ($\lsimeq 1,000$~G).  A field of a kilo-Gauss scales
under magnetic flux conservation to a neutron star field of $10^9$~G. Although
the peak of the low field distribution has still to be established
observationally, it is conceivable that the fields will be distributed in a
Gaussian manner about a peak that will map on to the observed field
distribution of the radio MSPs. Given the high mass transfer rates required
for AIC, the Ohmic diffusion time scale will be much larger than the accretion
time scale (Cumming 2002), so we expect the white dwarf field to be submerged
during the build up of the white dwarf mass prior to collapse. For the above
scenario to be viable, we need to postulate that the submerged field will
re-emerge without decay to its flux conserved value at the birth of the
neutron star. In this context, we note that there is no evidence of
accretion-induced field decay in the AM Herculis-type Cataclysmic Variables,
where a highly magnetic white dwarf has been accreting mass over billion years
from a companion. In fact, their well studied field configurations are very
similar to those observed and modelled in the isolated high field magnetic
white dwarfs (e.g. Wickramasinghe \& Ferrario 2000 and references therein).

We expect the white dwarf to be spun up to near break up velocity prior to
collapse (e.g. like the white dwarfs observed in dwarf novae).  However,
angular momentum (and mass) must necessarily be lost during the subsequent
collapse to a neutron star (Bailyn \& Grindlay 1990) so that detailed models
are required to establish the expected birth spin and mass distributions of
the resulting neutron star.  Dessart et al. (2006) have conducted
2.5-dimensional radiation-hydrodynamics simulations of the AIC of white dwarfs
to neutron stars.  Their calculations show that these lead to the formation of
neutron stars with rotational periods of a few (2.2-6.3) milliseconds. Hence,
even if the binary system were to be disrupted following a kick during the
AIC, these ``runaway'' newly born neutron stars would appear as isolated
radio-MSPs of the type currently observed.

A proportion of the white dwarfs that could be subjected to AIC will
inevitably belong to the high field group ($\sim 10^6-10^9$~G), and will
result in rapidly rotating (millisecond) pulsars with fields in the range
$10^{10}-10^{14}$~G on collapse, if we assume magnetic flux conservation (see
Figure 1 in FW).  This proportion could be as high as $50$\% because highly
magnetic white dwarfs tend to be more massive than their non-magnetic (or
weakly magnetic) counterparts (mean mass of $0.92\Msun$, Wickramasinghe \&
Ferrario 2005).  Assuming that the kicks are not field dependent and thus
preferentially disrupt these systems, we may also expect a group of MSPs with
high fields.  However, given the much higher spin-down rates of these objects,
and their low birth rates as compared to normal radio-pulsars, we expect them
to make a small contribution which would be dominated by the lowest field
objects in the distribution which would have the longest lifetimes as
radio-pulsars. A possible candidate could be the binary radio-pulsar
PSR~B0655+64 which has a relatively high magnetic field ($B=1.17^{10}$~G),
short orbital period ($P_{\rm orb}=1.03$~d) and is on a nearly circular orbit
(Damashek, Taylor\& Hulse 1978; Edwards \& Bailes 2001).

The population synthesis calculations of Hurley et al. (2002) yielded an AIC
rate that is two orders of magnitude higher than the LMXB(CC) rate that
results from the evolution of binary systems with primaries that are less
massive than $2\Msun$. A more detailed investigation of the AIC and core
collapse rates and orbital period distributions expected from such
calculations has been presented by Tout et al. (2007).  Here it is shown that
as with the core collapse route, \emph{the AIC route also generates binary
MSPs of all of the observed types}. We note in particular that a
class of long period ($P_{\rm orb} \ge 10$~d) binary MSPs with He white dwarf
companions is predicted, and this closely follows the observed
$P_{\rm orb}-M_{WD}$ relationship (Van Kerkwijk et al. 2005).

We conclude this section by noting that neutron stars that are formed via AIC
may also go through a mass transfer phase prior to their switching-on as radio
MSP.  We therefore expect that the known sample of LMXBs/IMXBs will have a
contribution from both neutron stars that have resulted from the core collapse
of massive stars and from the AICs. According to current estimates of the AIC
rates, the LMXBs(AIC)/IMXBs(AIC) may dominate over the
LMXBs(CC)/IMXBs(CC). However, because of field submersion, it may be difficult
at present to distinguish between these two possibilities. 

\section{Conclusions}

We have presented an analysis of the properties of the MSPs and placed
constraints on the magnetic field and the spin period distributions of the
MSPs at the time they turn on as radio emitters. We find that if we assume a
braking index $n=3$, the field distribution can be represented by a Gaussian
in the logarithm with mean $\log B({\rm G})=8.1$ and $\sigma_{\log B}=0.4$ and
the birth spin period by a Gaussian with mean $P_0=4$~ms and
$\sigma_P=1.3$~ms.  Our study, which allows for acceleration effects on the
observed spin-down rate, shows that (i) most MSPs are born with periods that
are close to the currently observed values (ii) the characteristic ages of
MSPs are typically much larger than their true age, and (iii) sub-millisecond
pulsars are rare or do not exist.  We also find a Galactic birth rate for the
MSPs of $\gsimeq 3.2\times 10^{-6}$ yr $^-1$.

We have used our results to discuss the relative merits of the
LMXB(CC)/IMXB(CC) and AIC scenarios that have been proposed for explaining the
origin of the MSPs.  Our main conclusions can be summarised as follows.

\begin {itemize}

\item[(a)]  The birth rate that we deduce for the MSPs is significantly larger (by a
factor $\sim 100$) than the semi-empirical birth rates quoted for the LMXBs/IMXBs
and is more in accord with the expected birth rates from the AIC route from
population synthesis calculations.

\item[(b)] The AIC scenario relates the MSP neutron star field distribution to
their white dwarf ancestry without invoking any kind of field decay, and finds
some support from the recently discovered bi-modality of the magnetic field
distribution of the white dwarfs.  The low fields of the MSPs may simply arise
from the low field component of the white dwarf magnetic field distribution
that is expected to peak below $10^3$~G and to scale to fields below $10^9$~G
under magnetic flux conservation. The nearly Gaussian distribution that we
deduce for the neutron star birth magnetic fields in the MSPs may thus  have a
more ready explanation on the AIC scenario.

\item[(c)] On the standard LMXB(CC)/IMXB(CC) picture, the majority of the MSPs
which have intermediate or long orbital periods, come from the IMXB(CC) route.
However, the predictions of the expected orbital period distributions from
population synthesis calculations are not in good agreement with the
observations of the MSPs.  It remains to be seen if better agreement with the
observed period distribution can be obtained with predictions from the AIC
route which is expected to also produce systems with the same range of orbital
periods and companion masses at the end of mass transfer prior to the
radio MSP phase.

\item[(d)] Population synthesis calculations predict that the standard
LMXB(CC) route results in a significant class of LMXBs with periods less than
$1$~hr, and these are observed as ultra compact LMXBs. However, there is not
as yet a single MSP with such a low period. A likely possibility is that the
mass of the companion has been reduced to negligible values by mass transfer,
or that the companion has been fully ablated by the time the pulsar turns
on. These systems could result in isolated MSPs.  However, the AIC route also
leads to similar systems and therefore end products.

\item[(e)] The peak in the spin period distribution of accretion and nuclear
powered X-ray pulsars occurs at $\sim 2$~ms, shorter that the $\sim 4$~ms that
we have deduced for the birth spin period of MSPs assuming a braking index of
$n=3$.  We find that the two distributions can be brought into closer
agreement if we assume a braking index $n=5$ which may suggest that the spin
down in the MSPs is dominated by angular momentum losses by gravitational
radiation or by magnetic multipolar radiation. Alternatively, this may be an
indicator that the two groups are not associated, and have neutron stars with
intrinsically different properties (e.g. mean mass) which is reflected in their
birth spin periods.

\end{itemize}

We differ a more detailed discussion of the expected outcomes from the AIC
route to a subsequent paper (Tout et al. 2007), where we present a detailed
comparison of the birth rates and orbital period distributions of the different
types of radio MSPs that result from the usual LMXB(CC)/IMXB(CC) route and the AIC
route.

We conclude by noting that if the AICs provide the dominant route leading to
the MSPs, one has to make the proposition that neutron star fields do not
decay, even in accreting stars, which remains contentious at the present time.
This issue is likely to be resolved by detailed observations, as has been done
in the case of white dwarfs where the general consensus appears to be that
there is no evidence for field decay either in single magnetic white dwarfs, or in
accreting magnetic white dwarfs in binaries.

\section*{Acknowledgements}

We thank Chris Tout and Jarrod Hurley for helpful discussions and the
anonymous Referee for a careful reading of our manuscript and for numerous
useful comments.

\end{document}